\documentclass[conference]{IEEEtran}
\usepackage{graphicx}

\usepackage{xcolor}

\usepackage{svg}

\usepackage{comment}

\ifCLASSINFOpdf
\else
\fi
\begin{document}
%
\title{Benchmarking Data Management\\ Systems for Microservices}

\author{\IEEEauthorblockN{Rodrigo Laigner}
\IEEEauthorblockA{
University of Copenhagen, Denmark\\
rnl@di.ku.dk}
\and
\IEEEauthorblockN{Yongluan Zhou}
\IEEEauthorblockA{
University of Copenhagen, Denmark\\
zhou@di.ku.dk}}


%


\maketitle



%
\IEEEpeerreviewmaketitle

\section{Introduction}
\label{sec:1}



Microservice architectures emerged as a popular architecture for designing scalable applications. This architecture promotes the decomposition of an application into independently deployable small services each encapsulating a private state~\cite{fowler:14}. Data exchanges and communication among microservices are often achieved via asynchronous events. This architecture enables practitioners to reap benefits associated with loose coupling, fault isolation, higher data availability, independent schema evolution, and increased scalability~\cite{LaignerKLSO20}.

While microservices have been extensively employed in industry settings for over a decade, existing microservice benchmarks fall short of reflecting core data management challenges that arise in practice~\cite{vldb2021}. As a result, it is difficult to advance data system technologies for supporting real-world, data-intensive microservice applications.

In this talk, we present \textit{Online Marketplace}, a microservice benchmark that incorporates data management challenges that existing benchmarks have not sufficiently addressed. These challenges include distributed transaction processing, data replication, consistent data and event querying and processing, and enforcing data integrity constraints. To start, we present the benchmark's application scenario, workload, and driver (Section~\ref{sec:benchmark}). Next, we prescribe data management criteria to match the aforementioned challenges and we explain different implementations of \textit{Online Marketplace} (Section~\ref{sec:impl}). We conclude by describing promising future work (Section~\ref{sec:conclusion}).
\section{Online Marketplace Benchmark}
\label{sec:benchmark}


Online marketplaces experience growing adoption over the globe~\cite{statista}. The workload incurred by interacting with clients as part of typical business transactions in these platforms often leads to a substantial amount of data being generated, processed, and stored. As a result, the complex functionalities and scaling requirements of these applications pose challenges to software teams in maintaining and evolving different modules independently, adapting to fluctuating workloads more holistically, isolating workloads from impacting other modules, and refraining failures from propagating to other microservices.



It is not a surprise many marketplace software providers have revealed the adoption of event-driven microservice architectures as a means to escape from aforementioned challenges~\cite{b2w}. These facts make online marketplaces a fit to inspire a microservice benchmark. Next, we summarize each microservice and its associated functionalities. 




\noindent\textbf{Application Scenario.} \textit{Cart} is responsible for managing individual cart instances for each customer. \textit{Product} encapsulates product information belonging to different sellers. \textit{Stock} manages inventory data. \textit{Order} contains key logic about the ordering process, including assigning invoice numbers, assembling the items with stock confirmed, and calculating order totals. \textit{Payment} is responsible for processing different payment methods and possible discounts, and confirming the order. Upon successful payment, the \textit{Shipment} creates shipment requests and puts items into packages. Eventually, shipped packages are delivered and their respective orders are deemed completed. \textit{Customer} is responsible for encapsulating customer data and updating customer statistics. \textit{Seller} is responsible for encapsulating seller data and serving seller-centric queries.

\noindent\textbf{Queries.} Five queries that are commonly used in the order processing lifecycle of marketplaces have been defined. Each aims to fulfill different types of clients, posing different characteristics.

\noindent\textbf{Customer Checkout.} After a series of cart operations (e.g., adding/deleting an item), a customer places an order by sending a checkout request. That triggers \textit{Cart} to assemble the items, apply updated prices (received from \textit{Product}) to items, and raise an event to trigger a new order processing. 

\noindent\textbf{Price Update.} A seller requests the price update of a given item to the \textit{Product} microservice. Upon processing the request, \textit{Product} generates an update for \textit{Cart}.

\noindent\textbf{Product Delete.} A seller requests the deletion of a product from the inventory. Upon processing the request, \textit{Product} also generates an event, so \textit{Stock} and \textit{Cart} can update their state accordingly.

\noindent\textbf{Update Delivery.} This transaction picks the first 10 sellers with undelivered packages in chronological order and sets their respective oldest order's packages as delivered. 

\noindent\textbf{Seller Dashboard.} Two queries are issued to compose the seller dashboard. The first is a continuous query that computes the financial amount of orders in progress by the seller, and the second returns the tuples used to compute the first.

\noindent\textbf{Driver.} We develop a benchmark driver in .NET to manage the experiment lifecycle. Its functionalities include data generation, data ingestion, system warm-up, submission of workload, statistics collection, and cleanup. In the talk, we intend to also cover practical challenges on workload submission including, but not limited to, accounting for deleted products while not impacting key distribution and providing safe concurrent accesses to data that form transaction inputs.

\noindent\textbf{Data Management Criteria.} To allow for proper comparison across data systems and platforms, we have also defined criteria for various data management issues that arise in microservice applications, some of which can be briefly summarized as follows. Business transactions crossing microservices, such as \textit{Customer Checkout}, must ensure all-or-nothing atomicity. We also define different correctness semantics for \textit{Product} replication to \textit{Cart}, including eventual and causal replication. Concerning data integrity constraints, we prescribe that stock items must always refer to existing products. 
For consistent data processing, the two queries that form the seller dashboard must reflect the same snapshot of the application state. Regarding event processing, events can be processed unordered or causally ordered. For instance, payment-related events must precede shipment events in the same order to meet causality.




\section{Implementations and Experiments}
\label{sec:impl}

To showcase \textit{Online Marketplace}, we implement four different versions of \textit{Online Marketplace} in two competing data platforms: Microsoft Orleans and Apache Statefun. Both platforms offer a runtime to execute distributed applications, programming models, and state management functionalities that aim to simplify microservice developments. 

\begin{figure}
\centering
\includegraphics[width=0.45\textwidth]{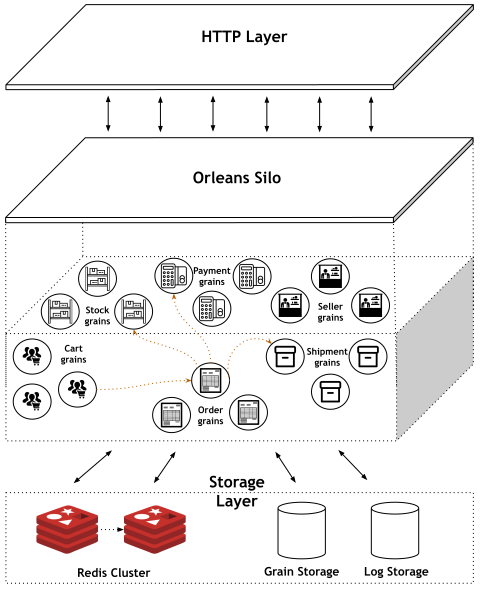}
\vspace{-3ex}
\caption{Customized Orleans-based Solution. \textit{HTTP Layer} parses HTTP requests and forwards them to the correct grains. \textit{Orleans Silo} provides location and life-cycle transparency for grains. Events are modeled as asynchronous messages exchanged by grains. \textit{Storage Layer} contains a primary-secondary deployment based on Redis to support causal replication of product updates, grain storage to manage grain states and log storage to store audit logging.}
\label{fig:custom_orleans}
\vspace{-4ex}
\end{figure}

We discuss issues encountered when implementing the benchmark in these competing platforms and the several shortcomings that prevented us from fulfilling the prescribed data management criteria. This matches recently reported results that found practitioners resort to weaving several heterogeneous systems together to fulfill their requirements~\cite{vldb2021}. Inspired by a common stack of technologies put forth by practitioners, we designed a full-featured implementation (Figure~\ref{fig:custom_orleans}), achieving all benchmark prescribed requirements. This shows \textit{Online Marketplace} can effectively pinpoint the core data management features pursued by practitioners.

We summarize the implementations and some results obtained under different workload settings. \textbf{Orleans Eventual.} It provides eventual consistency; that is, it does not ensure all actions are complete as part of a business transaction but exhibits the highest throughput.
\textbf{Orleans Transactions.} We use Orleans Transactions to implement ACID transactional guarantees to ensure all-or-nothing atomicity and concurrency control. However, this comes at a considerable overhead.
\textbf{Apache Flink Statefun.} Statefun is a dataflow-based platform that provides exactly-once processing. This implementation shows lower scalability compared to Orleans Eventual but outperforms Orleans Transactions by 2 times.
\textbf{Customized Orleans.} Solution based on Orleans Transactions that offloads consistent querying and causal replication to PostgreSQL and Redis, respectively. These features cannot be supported by the previous three settings. Our implementation introduces low overhead, hence its performance is comparable to Orleans transactions.

\section{Conclusion}
\label{sec:conclusion}



We have designed \textit{Online Marketplace}, a benchmark reflecting data management properties of real-world microservice deployments. Through our evaluation, we find no single data platform supports all the core data management requirements. This impedance leads practitioners to fall prey to complicated system stacks. We envision that the development of a data platform to support all data management features sought by practitioners can be fomented by \textit{Online Marketplace}. It is noteworthy our full experience can be accessed in~\cite{laigner2024benchmark}.








\bibliographystyle{IEEEtran}
\bibliography{main}

\end{document}